\documentclass[12pt,a4paper]{article}

\usepackage{graphicx}

\newcommand{\frat}[2]{\frac{\textstyle #1}{\textstyle #2}}
\newcommand{\vf}[1]{\mbox{\boldmath $#1$}}

\begin{document}
 \begin{center}
{\Large \bf   Gluon condensate behaviour at filling the Fermi sphere up}\\
 \vspace{0.5cm}A.E. Dorokhov$^1$, S.V. Molodtsov$^{1,2}$, G.M. Zinovjev$^3$
\\ \vspace{0.5cm} {\small $^1$Joint Institute for Nuclear Research,
RU-141980, Dubna, Moscow region, Russia}\\
 \vspace{0.5cm} {\small $^2$Institute of Theoretical and Experimental Physics,
 RU-117259, Moscow, Russia}
\\ \vspace{0.5cm} {\small $^3$N.N. Bogolyubov Institute for Theoretical Physics,
National Academy of Sciences of Ukraine, UA-03680, Kiev-143, Ukraine}
\end{center}
\vspace{0.5cm}

\begin{center}
\begin{tabular}{p{16cm}}
{\small{ The impact of filling up the Fermi sphere with the quarks, which dynamically
generated their masses on the instanton liquid at finite temperature
and baryonic/quark number density, is investigated. It is demonstrated, in particular,
that the boundary of chiral symmetry restoration phase transition is shifted to the
larger (about 100 MeV more) value of quark chemical potential compared to the
magnitude inherent in the Nambu-Jona-Lasinio model.}}
\end{tabular}
\end{center}
\vspace{0.5cm}

\section*{Introduction}
Impressive results obtained in experimental study of ultrarelativistic ion collisions at RHIC (Brookhaven)
and the experiments which are planned for the near future at
ALICE LHC (CERN) \cite{0} are standing in need of more accurate and precise theoretical predictions
for possible signatures of new states of strongly interacting matter with fastly growing acuity. However,
the theoretical advancement is much less appreciable especially in
the latest years. For example, the predictions of various approaches for the behaviour
of gluon condensate at finite $T$ and non-zero values of baryonic/quark chemical
potential $\mu$ which is a key quantity for theoretical analysis are still inconsistent
as before and at times simply conflicting. The possible changes appearing in the gluon
sector at such conditions and usually described by varying the constants of multiquark interactions as
the functions of $T$ and $\mu$ in the Nambu--Jona-Lasinio (NJL) model \cite{NJL},
need drawing almost inevitably the lattice numerical calculations of the gluon condensate \cite{rtw}
to be analysed. Practically to the same extent this remark is justified for the predictions of
the chiral perturbation theory (CHPT) \cite{chpt} and the QCD sum rules (SR) \cite{sr}. Both
approaches have rather limited reliability for results of calculations around the critical parameter
values. Actually, in  Ref. \cite{1} we have already tried to estimate the gluon condensate behaviour
in hot and dense medium using the instanton liquid (IL) model \cite{4}--\cite{IL2} as an operative tool.
In this case the screening impact of quarks filling the Fermi sphere up\footnote{Following \cite{PY}
we take this effect as the dominating one at high temperature though there exist the other interesting
possibilities \cite{Ripka}.} on the gluon condensate has been calculated for the massless quarks.

In this paper we consider the influence of quarks with the finite masses on the gluon condensate. In
the IL model the calculation of dynamical quark mass at zero temperature is grounded on making use
the zero mode approximation \cite{DPPob}. However, even that calculation runs into rather serious
technical difficulty (see also \cite{Sc}) while interpreting the loop quark diagrams at the chemical
potential values exceeding the magnitude of dynamical quark mass $\mu \ge M_q$. We are treating this
point here based on the NJL model and are not interested in the asymptotic large values of $\mu$
and $T$ (see, for example, \cite{son}) and omit an analysis of the colour superconducting phase
as well as the discussion of the difficulty in stabilizing the instanton ensemble which is rather often
resulted in the speculations about the 'realistic' structure of vaccum configurations.

\section{Approximating the vacuum configurations \\at finite $T$ and $\mu$}
Our purpose here is to find the practical and effective tool for evaluating the gluon condensate under
extreme conditions. Obviously, such a task has been pending for rather long time and the prevailing number of
scenarios to resolve it is grounded on the mean field approximation which supposes, actually, the
simplified description of a system. It will be a guiding element of our approach  while dealing
with the instanton ensemble. For example, vacuum correlation function $\langle A_\mu(x)A_\nu(y)\rangle$
of mean field description is transformed into the correlator which in the context of our approach leads
to the mass generation for gluon field and, hence, to the colour screening factor.

As well known in the IL model at zero values of $\mu$ and $T$ the superposition of (anti-)instantons
in the singular gauge 
\begin{equation}
\label{1}
A^a_{\mu}(x;\gamma)=\frat2g~\omega^{ab}\bar\eta_{b\mu\nu}~a_\nu(y)~,~~~
a_\nu(y)= \frat{\rho^2}{y^2+\rho^2}~\frat{y_\nu}{y^2}~,~~~y=x-z~,~~\mu,\nu=1,2,3,4~,
\end{equation}
(where $\rho$ is a pseudo-particle size, $\omega$ is a matrix of its colour orientation and $z$ is its center
coordinate) is considered as the ground vacuum field saturating the QCD generating functional (dealing
with anti-instantons one has to change the't Hooft symbol $\bar\eta \to \eta$).
The  QCD generating functional is evaluated to be as
\begin{equation}
\label{2}
 Y=\sum_{N=1}^\infty\frat{1}{N!}~\prod_{i=1}^N~\int d\gamma_i~d_0(\rho_i)~e^{-\beta~U_{int}(\gamma)}
 =\sum_{N=1}^\infty
\frat{1}{N!}~\prod_{i=1}^N~\int d\gamma_i~e^{-E(\gamma)}~,
\end{equation}
$$E(\gamma)=\beta~U_{int}(\gamma)-\sum~\ln d_0(\rho_i)~,~~~\gamma=(z,\rho,\omega)~,$$
here
\begin{equation}
\label{3}
d_0(\rho)=\frat{1}{\rho^5}~\widetilde\beta^{2N_c}~e^{-\beta(\rho)}~
\end{equation}
is the distribution function over the size of individual instanton (dilute instanton gas approximation)
\cite{4}, $d\gamma_i=d^4z_i~d\omega_i~d\rho_i$ is the integration element,
$$\beta(\rho)=\frat{8\pi^2}{g^2}=-b~\ln (C_{N_c}^{1/b} \Lambda\rho)$$
is the single instanton action
($\Lambda=\Lambda_{\overline{MS}}=0.92 \Lambda_{P.V.}$) with the constant $C_{N_c}$
depending on the renormalization scheme $C_{N_c}\approx\frat{4.66~\exp(-1.68 N_c)}
{\pi^2 (N_c-1)!(N_c-2)!}$ with another parameter $b=\frat{11~N_c-2~N_f}{3}$.
The auxiliary coefficients
$\widetilde \beta=-b~\ln(\Lambda \bar\rho)$ ш $\beta$ in the exponent of Eq. (\ref{2}) are
fixed at the characteristic scale $\bar\rho$ (pseudo-particle average size). Assuming the
topologically neutral instanton liquid we do not differ the instantons and anti-instantons
and $N$ denotes (when used) the total number of pseudo-particles which occupy the volume $V$.

Taking into account the interaction of instantons with vacuum fluctuations is effectively
presented by appearance of the screening factor in the distribution (\ref{3})
\begin{equation}
\label{d}
d(\rho)=\frat{1}{\rho^5}~\widetilde\beta^{2N_c}~e^{-\beta(\rho)-\zeta\rho^2}~,
\end{equation}
where the magnitude of screening coefficient $\zeta$ is dependent on the choice of
superposition ansatz. For the pseudo-particles in the singular gauge the interaction term taken
in the pair interaction approximation is \cite{IL2}
$$\int
d\omega_1~d\omega_2~dz_1~dz_2~U_{int}(\gamma_1,\gamma_2)=V~\xi^2~\rho_1^{2}~\rho_2^{2}~,$$
with the constant $\xi^2=\frat{27~\pi^2}{4}\frat{N_c}{N_c^{2}-1}$.
It is interesting to notice here that the configurations used in the valley method \cite{by}
result in the significantly smaller value (about one order) of coefficient $\xi$ \cite{yvf}.
Besides, the screening factor can be steadily extracted from the lattice data as
$\lambda_A\sim 0.22$ fm \cite{dgiac} with the configuration cooling procedure. The corresponding
configurations are reasonably well fitted by the instanton ensemble as was shown \cite{fit}, although
the analysis of optimal instanton configurations in the mean field approximation is worthy
of special study and will be done in the separate paper \cite{wenew}.

The convexity property of exponential function allows us to estimate the partial contribution into
the generating functional Eq. (\ref{2}) at each value of $N$ by the following approximating form
\begin{equation}
\label{4} Y\ge Y_{approx}=Y_1~\exp(-\langle E-E_1\rangle)~,
\end{equation}
which can be presented \cite{IL2} as
\begin{equation}
\label{5}
Y_{approx}=e^{-X}~,~~X=N\left(\frat{\nu}{2}+1\right)~[\ln(n/\Lambda^4)-1]-
N\ln \left[C_{N_c}\widetilde\beta^{2N_c}(\beta\xi^2\nu)^{-\nu/2}\frat{\Gamma(\nu)}{2}\right]~,
\end{equation}
where $n=N/V$, $\nu=(b-4)/2$.
Then the respective parameters of IL are defined by the maximum in $n$ of generating functional
with the interrelation of instanton average size and its density taken into account
\begin{equation}
\label{6}
\frat{\nu}{\overline{\rho^2}}=\beta \xi^2 n \overline{\rho^2}~.
\end{equation}

Now calculating the maximum of $X$ in $n$ we have to resolve the following equation
\begin{equation}
\label{8b}
-\left(\frat{\nu}{2}+1\right)~\ln(n/\Lambda^4)+
\ln \left[C_{N_c}\widetilde\beta^{2N_c}(\beta\xi^2\nu)^{-\nu/2}\frat{\Gamma(\nu)}{2}\right]
+n~\frat{2N_c}{\widetilde\beta}\frat{d\widetilde\beta}{dn}-
n~\frat{\nu}{2\beta}\frat{d\beta}{dn}=0~.
\end{equation}

Owing to the relation (\ref{6}) we have
$$\frat{1}{\beta}\frat{d\beta}{d\bar\rho}+\frat{1}{n}\frat{d n}{d\bar\rho}+\frat{4}{\bar\rho}=0~.$$
From the other side
$\frat{d\beta}{d\bar\rho}=-\frat{b}{\bar\rho}$, $\frat{d\widetilde\beta}{d\bar\rho}=
\frat{d\beta}{d\bar\rho}$.
Rewriting the derivative of $\beta$ in the density as
$\frat{d\beta}{dn}=\frat{d\beta}{d\bar\rho}/
\frat{d n}{d\bar\rho}$,
we come to the system of equations
\begin{equation}
\label{9b}
\frat{d\beta}{dn}=\frat{1}{n}~\frat{b~\beta}{4\beta-b}~,
~~\frat{d\widetilde\beta}{d n}=\frat{d\beta}{dn}~.
\end{equation}
Finally resolving the system of transcendental equations we can determine the equilibrium
IL parameters.

At the finite temperature the configuration saturating the generating functional is changed by
the superposition of (anti-)colorons \cite{DM88}, \cite{novak} which are the periodical in
the Euclidean 'time' (with the period of $T^{-1}$) solutions of the Yang-Mills equations \cite{5} i.e.
\begin{eqnarray}
\label{7}
A^{a}_\mu(x,\gamma,T)&=&-\frat{1}{g}~\omega^{ab}~\bar\eta_{b\mu\nu}~
\partial_\nu \ln \Phi(x,T),~\nonumber\\[-.2cm]
\\[-.25cm]
\Phi(x,T)&=&1+\frat{\pi \rho^2 T}{r}\frat{\sinh(2\pi r T)}
{\cosh(2\pi r T)-\cos(2\pi\tau T)}~. \nonumber
\end{eqnarray}
Here $r=|{\vf x}-{\vf z}|$ defines the distance from the coloron center $z$ in three-dimensional
space, $\tau=x_4-z_4$ is the 'time' interval. It can be easily seen that the solution is transformed
into the (anti-)instanton one in the singular gauge at temperature going to zero. Clearly, the
distribution function over the coloron size \cite{PY}, \cite{scm} is also changed
\begin{equation}
\label{8}
d(\rho;\mu,T)=d(\rho)~e^{-\eta^{2}(\mu,T)~\rho^2}~,~~\eta^2(\mu,T)=2~\pi^2~
\left[T^2~\frat{N_c}{3}+\sum_{f=1}^{N_f}\Pi^{f}(\mu,T)\right]~.
\end{equation}
The first term of the screening factor describes the one-loop gluon contribution into the effective
action and the second term generated by quark contribution in one-loop approximation can be exactly
calculated and is free of the 'bad' singularities \cite{ap}. The 'time' component of polarization
tensor generated by quark of fixed colour has the form
\begin{eqnarray}
\label{16a}
\Pi^{f}_{44}(k_4,\omega)&=&\frat{k^2}{\pi^2\omega^2}~\int_0^{\infty}
\frat{dp~p^2}{\varepsilon_p} ~n_p
\left[1+\frat{4\varepsilon_p^{2}-k^2}{8pk}\ln\frat{(k^2+2p\omega)^2+4\varepsilon_p^{2}k_4^{2}}
{(k^2-2p\omega)^2+4\varepsilon_p^{2}k_4^{2}}-\right.\nonumber\\
&-&\left.\frat{\varepsilon_pk_4}{p\omega}~\arctan \frat{8
p\omega~\varepsilon_p k_4}
{4\varepsilon_p^{2}k_4^{2}-4p^2\omega^2+k^4}\right]~,\nonumber
\end{eqnarray}
here $\omega=|{\vf{k}}|$, $k^2=\omega^2+k_4^{2}$, $\varepsilon_p=(M_{q}^2+{\vf{p}}^2)^{1/2}$
where $M_{q}$ is the quark mass, $n_p=n_p^{-}+n_p^{+}$, $n_p^{-}=(e^\frac{\varepsilon_p-\mu}{T}+1)^{-1}$,
$n_p^{+}=(e^\frac{\varepsilon_p+\mu}{T}+1)^{-1}$ ($n_p^{-}, n_p^{+}$ are the densities of anti-quarks
and quarks, respectively). When summed up over all the components the polarization tensor can be
presented in the following form
\begin{equation}
\label{17a}
\Pi^{f}_{\mu\mu}(k_4,\omega)=\frat{2}{\pi^2}~\int_0^{\infty} \frat{dp~p^2}{\varepsilon_p}~ n_p
\left[1+\frat{2 M_{q}^2-k^2}{8pk}\ln\frat{(k^2+2p\omega)^2+4\varepsilon_p^{2}k_4^{2}}
{(k^2-2p\omega)^2+4\varepsilon_p^{2}k_4^{2}}\right]~.
\end{equation}
It is clear when the zero-component $k_4=0$ the dominant contribution into the gluon mass at
small values of $\omega$ comes from the first term (a unit) and the space components are negligible.
In particular, at $\omega=0$ it will be
\begin{equation}
\label{18a}
\Pi^{f}(\mu,T)=\Pi^{f}_{44}(0,0)=\frat{2}{\pi^2}~\int_0^{\infty} \frat{dp~p^2}{\varepsilon_p}~n_p~.
\end{equation}
Then at $T=0$ we have
$\Pi^{f}(\mu,0)=\left[\frat{(\mu^2-M_{q}^2)^{1/2}\mu}{\pi^2}-
\frat{M_{q}^2}{\pi^2}\ln\frat{\mu+(\mu^2-M_{q}^2)^{1/2}}{M_{q}}\right]$.
In order to calculate the IL equilibrium parameters as the functions of $\mu$ and $T$ one has to
minimize the approximating functional (\ref{5}) making the substitutions of (\ref{6}) and (\ref{9b})
for
\begin{equation}
\label{21b}
\frat{\nu}{\overline{\rho^2}}=\eta^2+\beta \xi^2 n \overline{\rho^2}~,
\end{equation}
and
\begin{equation}
\label{22b}
\frat{n}{\beta}\frat{d\beta}{dn}=\frat{b}{4\beta-b+\frac{2\eta^2\bar\rho^2\beta}{\nu-\eta^2\bar\rho^2}}~.
\end{equation}
correspondingly.

\section{Quark mass generation in stochastic field}
It is anticipated that in the IL model the quarks are considered to 'live' (and to be influenced)
in the stochastic (anti-)instanton ensemble which is defined by the following generating functional
\begin{equation}
\label{10}
Z=\int D[\psi] D[\bar\psi]~\langle e^{\int dx ~{\cal L}_q}\rangle_A~,~~~
{\cal L}_q=\bar\psi(x)\left(i\hat\partial_x+\sum^N_{k=1} g\hat A(x;\gamma_k)\right)\psi(x)~,
\end{equation}
where the averaging over (anti-)instanton ensemble is implied. The consistency requirement
for effective Lagrangian in the Hartree approximation results in the equation for the quark
Green function \cite{sim} which reads as
\begin{equation}
\label{12} M(p)=-\frat{N}{V}~\frat{1}{N_c}\sum_{n=2}^\infty \int
\prod_{k=1}^n \frat{d q_k}{(2\pi)^4}~
(2\pi)^4~\delta^4\left(\sum_{i=1}^n q_i\right)~Tr\left( g~\hat
A(q_1)~S(p-q_1)\dots g~\hat A(q_n)\right)~.
\end{equation}
Being summed up the right hand side of Eq. (\ref{12}) can be presented in the compact form
as \cite{pob}
\begin{equation}
\label{13} M(p)=\frat{1}{N_c V}~ Tr\sum_{i=1}^N ~\langle
p|~\left[S- (g~ \hat A(q_i))^{-1}\right]^{-1} |p\rangle~,
\end{equation}
(in such a form the averaging over the pseudo-particle location $z$ and calculation of
colour trace is meant). Analyzing the solution in the form
\begin{equation}
\label{11}
S(p)=\frat{ 1}{\hat p -i M(p)}~,
\end{equation}
where M(p) denotes the quark mass, one can calculate the highest term of expansion in the IL
density (presented by the zero quark mode $\Phi(p)$ in the instanton field) as \cite{pob}
$$M(p)\sim n^{1/2}~\frat{p^2~\Phi^2(p)}{\left[\int
\frat{dp}{(2\pi)^4}~p^2~\Phi^4(p)\right]^{1/2}}~.$$
At finite quark chemical potential the derivative $i\hat\partial$ in Eq. (\ref{10}) should be
substituted for $i\hat\partial-i\hat\mu$ where $\hat\mu=\mu\gamma_4$. Then quark Green function
(\ref{11}) develops the following form
\begin{equation}
\label{14}
S(p;\mu)=\frat{ 1}{\hat p +i\hat\mu-i M(p;\mu)}~.
\end{equation}
where
$$M(p;\mu)\sim n^{1/2}~\frat{(p+i\mu)^2~\Phi^2(p;\mu)}{\left[\int
\frat{dp}{(2\pi)^4}~(p+i\mu)^2~\Phi^2(p;\mu)~\Phi^2(p;\mu)\right]^{1/2}}~,$$
and the overt expression of the zero mode could be found, for example, in \cite{dc}. With the
chemical potential increasing and reaching the values of dynamical quark mass order
($\mu\sim M_q$) the quark mass magnitude $M(p;\mu)$ begins to increase as a power. The similar
situation with the dynamical mass increase takes place for the approach in which the
unperturbated quark Green function $S_0$ is approximated by the zero modes \cite{dc},
\cite{wesc}{\footnote{Determining the saddle point parameter in this case one encounters the problem of
calculating a loop integral in which the pole of Eq. (\ref{14}) appears on the real axis. The
treatment of that integral as a principle value gives its real part only which is not enough, of
course. However, this problem is softened by the situation that the pole appearance on the real
axis occurs in the local vicinity of transition point into the colour superconducting phase and,
therefore, more
precise definition of the saddle point parameter looks superfluous. However, the general problem of
calculating the loop integrals is entirely hot and actual in the context of analyzing the chiral
symmetry restoration. (Prof. T. Hatsuda has drawn our attention to this aspect of the problem and
the authors are very grateful to him for that.)}}. Such a behaviour is non-physical and contradicts
to the intuitive expectations. It seems, the situation could be improved by taking into account the
non-zero mode contributions but very complicated analytical structure of the corresponding expressions
makes this calculation practicaly hopeless. Thus, the question about the estimate of non-zero mode
contribution is still vague (see, however, \cite{dkz}). The proposition to treat poles as in
Ref. \cite{Sc} leads, unfortunately, to unphysically small values of the chemical potential of chiral
symmetry restoration phase transition. It is interesting to notice here that the zero mode
approximation is quite reliable even at the finite temperature if one confines oneself to work with
the chemical potential values not larger than the dynamical quark mass $\mu\sim 300$ MeV. The quark
condensate estimates are quite suitable in this case even if one makes use simply non-coloron
zero mode. These results put forward the task of searching the effective approximations for the
equations of type (\ref{12}).
The significant progress in studying the systems of quarks at finite temperature and chemical
potential has been reached in the framework of NJL approach. (Let us remember here that at finite
temperature the integration over the fourth component in Eq. (\ref{10}) should be performed in the
interval from zero till $T^{-1}$ and gluon fields obey the periodic boundary conditions whereas
the fermion fields obey the anti-periodic ones.)

As it is difficult to handle (anti-)instanton ensemble directly we are going to retain some
essential features of (anti-)instanton configuration contribution and approximate it with the
simplest form. Actually, we suppose the existence of superposition of stochastic randomly
oriented color gluon fields in the Euclidean space. These fields have the $\delta$-function
form with their randomly distributed centers $z$, i.e.
\begin{equation}
\label{15}  A_\mu(x)= U^\dagger \tau^a U~ a^a_{\mu}
~(2\pi)^4~\delta^{(4)}(x-z)~,~~A_\mu(p)=~ U^\dagger \tau^a U~
a^a_{\mu} ~(2\pi)^4~e^{ipz}= A_\mu~~e^{ipz}.
\end{equation}
It is clear if one considers one-particle correlations only (just
what is done for the pseudo-particle ensemble) the simplest
non-trivial correlation function $\langle \hat A(x_1)\hat
A(x_2)\rangle_{z,U}$ will lead to the point-like interaction of
quarks $$\langle \hat A(x_1)\hat
A(x_2)\rangle_{z,U}\sim\delta^{(4)}(x_1-x_2)~$$ which is specific
for NJL. In further analysis we do not need to know the concrete
form of stochastic factor $\hat A$ and do not specify it here.
Searching the solution of Eq. (\ref{12}) in the form $M(p)=M$ we
introduce the auxiliary function $\psi(q,p)$ the following way
\begin{eqnarray}
\label{16} &&M=~\frat{-i~n}{N_c}~ \int \frat{d
q}{(2\pi)^4}~Tr~\langle \hat A(q-p)~\frat{1}{\hat q +i\hat\mu -iM}~
\psi(q,p)\rangle_{z,U}~,\nonumber\\[-.2cm]
\\[-.25cm]
&&\psi(p,p')=\hat A(p-p')+ \int \frat{d q}{(2\pi)^4}~\hat
A(q-p)~\frat{1}{\hat q +i\hat\mu -iM}~ \psi(q,p')~,\nonumber
\end{eqnarray}
(of course, we imply non-zero quark chemical potential). Presenting the solution for $\psi(q,p)$
in the form $\psi(q,p)=\psi~e^{i(q-p)z}$ we are able to obtain the following equation to determine
the function $\psi$ of our interest
\footnote{Another utmost regime where the correlation length is supposed to be infinitely large
$\langle A(x)A(y)\rangle=A^2$ is also very interesting. It was analyzed in the Keldysh model
\cite{k} and the exact solution was found out. The complete summation of series for the quark
Green function results in the expression as
$$S(p)=\frat{1}{(2\pi)^{1/2}}~\int_{-\infty}^{\infty}dt~e^{-\frat{t^2}{2}}~\frat{1}{\hat p+\hat\mu-t~A}~,$$
(it is given here in the Minkowski space). Apparently, it has no poles similar to the
non-relativistic Green function as well as the 'analytical' model of confinement \cite{n}.}
\begin{equation}
\label{17} \psi=\hat A+\hat A~\int \frat{d q}{(2\pi)^4}~\frat{1}{\hat q +i\hat\mu -iM}~\psi~.
\end{equation}
As in the NJL model Eq. (\ref{17}) requires the regularization. Here we are using the conventional
procedure of three-dimensional momentum regularization \cite{NJL} which allows us to obtain
\begin{eqnarray}
\label{18} &&I=\int \frat{d q}{(2\pi)^4}~\frat{1}{\hat q +i\hat\mu
-iM}=iC~\gamma_4+iDM~,\\
&&C=-\frat{\theta(\mu-M)}{(2\pi)^2}\frat{(\mu^2-M^2)^{3/2}}{3}~,\nonumber\\
&&\parbox[b]{6.in}{$ D=
 \left \{ \begin{array}{l}
\frat{1}{8\pi^2} \left[\widetilde\Lambda
~(\widetilde\Lambda^2+M^2)^{1/2} -M^2\ln
\left|\frat{\widetilde\Lambda + (\widetilde
\Lambda^2+M^2)^{1/2}}{M}\right|\right]~,~~\mu\leq M  \\[-.1cm]
\\[-.1cm]
\frat{1}{8\pi^2} \left[\widetilde\Lambda~ (\widetilde
\Lambda^2+M^2)^{1/2} -\mu~ (\mu^2-M^2)^{1/2}-M^2 \ln
\left|\frat{\widetilde\Lambda + (\widetilde
\Lambda^2+M^2)^{1/2}}{\mu+(\mu^2-M^2)^{1/2}}\right|\right]~,~~\mu > M
\end{array} \right.$}\nonumber
\end{eqnarray}
$\widetilde\Lambda$ denotes here the cut-off value of three dimensional momentum in the $I$ integral.
Finally, we have for the solution of Eq. (\ref{17}) the following result
\begin{equation}
\label{19} \psi =\frat{B+C^2A^2}{B(1+D^2A^2M^2)}~(\hat A +iDA^2 M)+
\frat{iC}{B(1+D^2A^2M^2)}~(\hat A +iDA^2 M)~\gamma_4~(\hat A
+iDA^2 M)~,
\end{equation}
where $B=1-2iC A_4 -C^2A^2+D^2A^2M^2$.
Using this solution in Eq. (\ref{16}), averaging over the colour orientation and holding the highest
terms of the $N_c$ expansion we come to the mass gap equation
\begin{equation}
\label{20} M = 4 n~\frat{DA^2M}{1+D^2A^2M^2-C^2A^2}~.
\end{equation}
It enables to formulate the condition which signals the breakdown of chiral symmetry (the
generation of quark mass) if such a constraint is obeyed
\begin{equation}
\label{21} A^2 > \frat{1}{4 n~D-D^2M^2+C^2}~.
\end{equation}
If one neglects the contributions proportional to $A^2$ in the denominator of Eq. (\ref{20})
the gap equation of the NJL model with the coupling constant $G$ of four-fermion interaction
is exactly reproduced
\begin{equation}
\label{22}
G=4n~A^2~\frat{\widetilde\Lambda^2}{8\pi^2}~.
\end{equation}
In order to receive the qualitative estimates it is worthwhile using the characteristic cutoff
parameter of the NJL phenomenology $\widetilde\Lambda\sim 600$ MeV. In our estimates we rely on
the constraints $D^2M^2A^2\ll 1$,  $C^2 A^2 \ll 1$ only.

\begin{figure*}[!tbh]
\begin{center}
\includegraphics[width=0.5\textwidth]{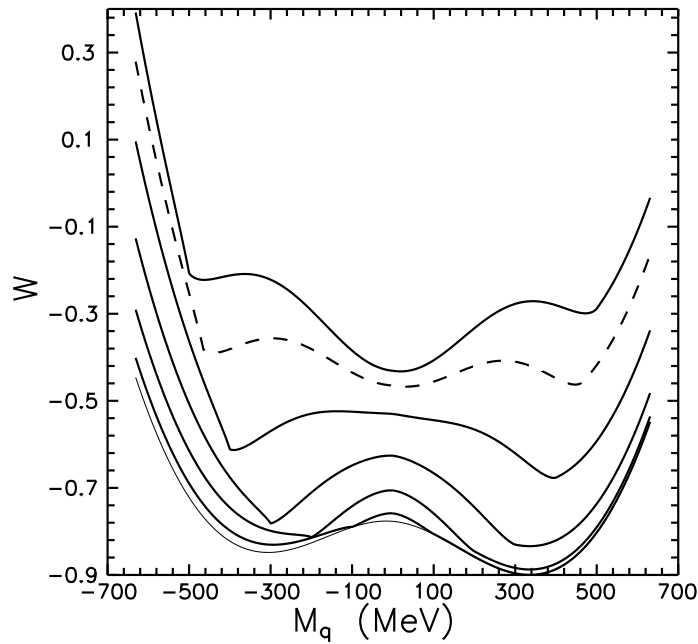}
\end{center}
  \vspace{-7mm}
 \caption{The effective potential $W$ as the function of quark mass for different values
of chemical potential  $\mu=0$ (lower curve), $\mu=100$,
$\mu=200$, $\mu=300$, $\mu=400$ $\mu=450$, $\mu=500$ MeV.}
  \label{f1}
\end{figure*}

The estimate of $A^2$ for instanton ensemble could be obtained
from the corresponding correlation function \cite{we} $$\langle
A(x-z)~A(y-z)\rangle_{z,U}=F(x-y)~.$$ Actually, we have
$F(0)=\frat{4\pi^2}{N_c^{2}-1}~\rho^2$, which means
$A_x^{2}\sim\frat{1}{\rho^2}$ and then we receive for the Fourier
component $A_p^{2}\sim \rho^6$. Taking into account that $\rho\sim
\widetilde\Lambda^{-1}$ and $D\sim\widetilde\Lambda^2$ (see Eq.
(\ref{18})) we obtain $D^2 A^2 M^2\sim
\frat{M^2}{\widetilde\Lambda^2}$. The standard parameter values of
the NJL model provide us with the small magnitude of this factor
and it could be neglected. Surely, it is a fairly serious argument
in favour of using the developed approach.

\section{Approximation of NJL and IL model}
Our above analysis demonstrates that approximating the instanton correlator with the delta-
function form and using the regularized NJL model at the same time is the fully compatible
procedure. Moreover, Eq. (\ref{22}) provides us with the possibility to consider the
interrelation of gluon and quark sectors. It relates the constant $G$ and the IL parameters
such as the IL density $n$ and average potential $A$ and $G$ is related to the dynamical
quark mass $M=M(0)$ which defines the screening effect.

Let us now remind  that the generating functional of the NJL model has the following form \cite{NJL}
\begin{eqnarray}
\label{23}
&&Z=e^{-\Omega}~,\nonumber\\[-.2cm]
\\[-.25cm]
&&\Omega=G_0 \sigma^2 -\frat{N_f N_c}{\pi^2}
\int_0^{\widetilde\Lambda} p^2 \varepsilon_p dp - \frat{N_f
N_c}{\pi^2}T\int_0^{\widetilde\Lambda} p^2 \left[ \ln
\left(1+e^{-\frat{\varepsilon_p+\mu}{T}}\right)+ \ln
\left(1+e^{-\frat{\varepsilon_p-\mu}{T}}\right)\right]dp,\nonumber
\end{eqnarray}
where $\varepsilon_p=(M^2+{\vf{p}}^2)^{1/2}$, $M=m-2G_0~ \sigma$, $m$ is the current quark mass and
for the quark condensate we have
\begin{equation}
\label{24} \langle \bar\psi \psi \rangle =-M~\frat{
N_c}{\pi^2} \int_0^{\widetilde\Lambda}
\frat{p^2}{\varepsilon_p}~(1-n_p^{-}-n_p^{+})~dp~,\ \ \ \
\sigma=N_f\langle \bar\psi \psi \rangle
\end{equation}

\begin{figure*}[!tbh]
\begin{center}
\includegraphics[width=0.5\textwidth]{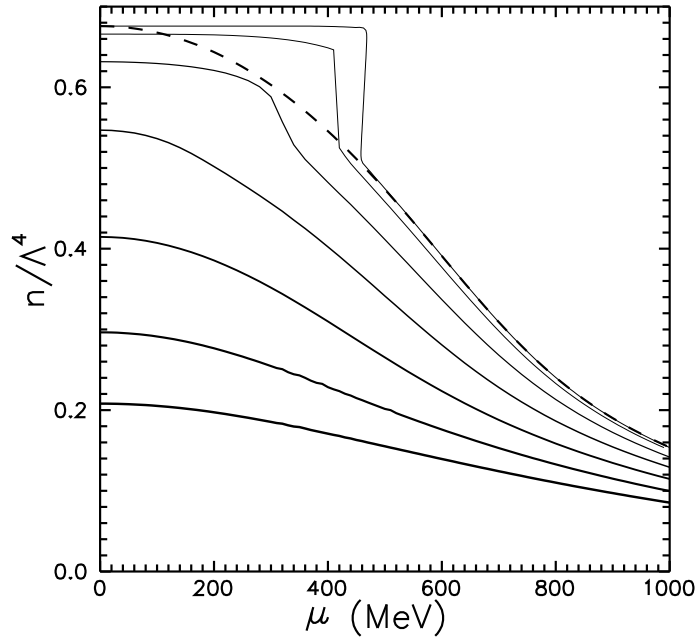}
\end{center}
  \vspace{-7mm}
 \caption{The IL density as the function of chemical potential for different values of temperature
(the IL density is decreasing with the temperature increasing)
$T=0$, $T=50$, $T=100$, $T=150$, $T=200$, $T=250$, $T=300$ MeV.}
  \label{f2}
\end{figure*}

\noindent
The quark mass is defined  by calculating the minimum of $\Omega$ as the function of $M$
(or quark condensate $\sigma$) and the coupling constant $G_0$ together with the cutoff
parameter $\widetilde\Lambda$ is fixed phenomenologically (by fitting the experimental data).
We suppose to take this quantity as an estimate of quark determinant while quarks are in
the stochastic field of (anti-)instantons and modify the determinant aiming to include
the interrelation of quark and gluon sectors. In this way we use (instead of $G$) in
Eq. (\ref{23}) $G\to\frat{n}{n_0}~G_0$ where $n_0$, $G_0$ are the IL density and the constant
of fourquark interaction at zero temperature
and zero chemical potential. In full analogy with the IL model it is easy to understand
the parameter $\hat A$  should generate the factor of the $\rho^3$ type and the substitution
of the coupling constant for $\frat{n\rho^6}{n_0\rho_0^{6}}~G_0$ looks quite natural. On the
other hand as the simplest option we could take the  cutoff parameter in the quark sector
$\widetilde\Lambda$ unchanging as the parameter $\hat A\sim\widetilde\Lambda^{-3}$. Thus,
for the quarks in stochastic instanton ensemble we should change $\Omega$ in Eq. (\ref{23}) for
\begin{equation}
\label{25} \Omega=\frat{n}{n_0}G_0 \sigma^2 -\frat{N_f N_c}{\pi^2}
\int_0^{\widetilde\Lambda} p^2 \varepsilon_p dp - \frat{N_f
N_c}{\pi^2}T\int_0^{\widetilde\Lambda} p^2 \left[ \ln
\left(1+e^{-\frat{\varepsilon_p+\mu}{T}}\right)+ \ln
\left(1+e^{-\frat{\varepsilon_p-\mu}{T}}\right)\right]dp.
\end{equation}
Apparently, the vacuum parameters of the IL model and the NJL one should not change. In order
to realize that one should make the corresponding subtractions just to retain the effect caused by
the quarks filling the Fermi sphere up because the interrelation of vacuum (at zero $T$ and $\mu$)
quark and gluon fields has been already discounted effectively in the running coupling constant and
by tuning the NJL model parameters. The (anti-)instanton ensemble and quark field are described by
the product of functionls $Y_{approx}$ and $Z$. Therefore, for the effective potential we have
$$W=X+\Omega~.$$

\begin{figure*}[!tbh]
\begin{center}
\includegraphics[width=0.5\textwidth]{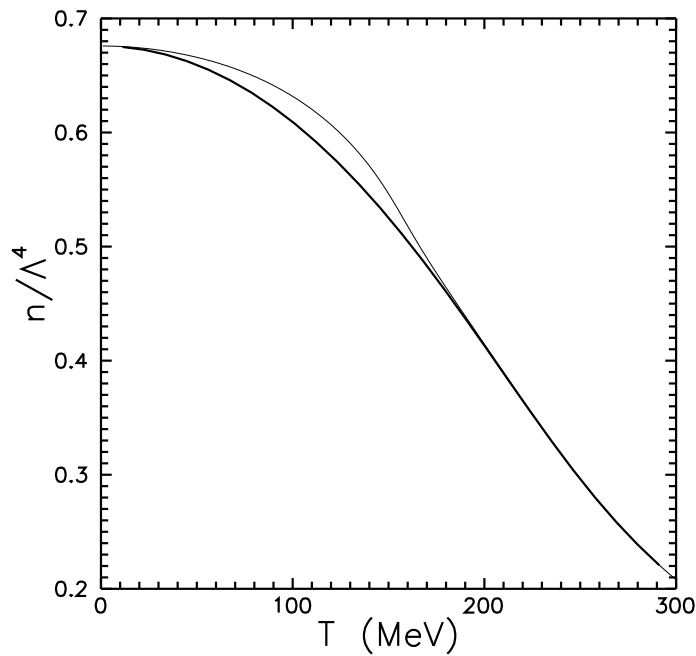}
\end{center}
  \vspace{-7mm}
 \caption{The IL density as the function of the temperature $T$. The lower curve corresponds to
the calculation with massless quarks.}
  \label{f3}
\end{figure*}

The equilibriumn IL parameters are defined by the effective
potential minimum on the IL density, i.e. by $\frat{\partial
W}{\partial n}=0$. However, as was declared  the following
subtraction
\begin{equation}
\label{26} \frat{\partial X}{\partial n}+\frat{\partial
\Omega}{\partial n}-\frat{\partial \Omega}{\partial n}
\left|_{\mu=0, T=0}\right.=0~.
\end{equation}
should be done. Similar operation should be executed at determining the quark mass
$\frat{\partial W}{\partial M}=0$, i.e.
\begin{equation}
\label{27} \frat{\partial X}{\partial \eta^2}\frat{\partial\eta^2}{\partial M} -
\frat{\partial X}{\partial \eta^2}\frat{\partial\eta^2}{\partial M}\left|_{\mu=0,
T=0}\right. +\frat{\partial \Omega}{\partial M} =0~.
\end{equation}
Two first terms of Eq. (\ref{27}) are the result of the fact that the overt dependence on the quark
mass in the contribution into the effective potential is available in the screening factor $\eta^2$
only. Strictly speaking one should integrate till the momentum order of $\widetilde\Lambda$ in
Eq. (\ref{18a}), too. However, such an amplification is superfluous as the detailed analysis shows.
In practice, obviously, it is simpler not to resolve the transcendental equation (which has two
branches, at least) but calculate simply the minimum of effective potential in $M$. It is easy to
understand that for the concrete form of our quark effective potential $\Omega$ the equation for
determining the equilibrium IL parameters coincides with the vacuum one because the direct
dependence on the IL density $n$ is present in the first term of Eq. (\ref{25}) only. Just because
of that reason two last terms of Eq. (\ref{26}) are canceled. The subtraction in the second  equation
of (\ref{27}) should not be performed as $\frat{\partial X}{\partial \eta^2}=-\frat{n\bar\rho^2}{2}$
and the function $\eta^2=0$ at $\mu < M$. Thus, the equilibrium IL parameters are defined by the
same scheme as before \cite{1} and minimum of the generating functional $W$ in $M$ fixes the dynamical
quark mass.

Here we are using the following set of the NJL model parameters \cite{NJL} (T. Hatsuda, T. Kunihiro).
We take for the current
mass of $u$ and $d$ quarks the same value $m=5.5$ MeV, for the cutoff parameter as
$\widetilde\Lambda=631$ MeV and for the ratio of the coupling constant to its critical value as
$\alpha=G_0/G_c=1.33/N_f$, $\left(G_c=\frat{\pi^2}{N_c\widetilde\Lambda^2}\right)$. Such a set of
parameters results in the following values of the $\pi$-meson mass $m_\pi=139$ MeV and the
constant of pion decay $F_\pi=93$ MeV.

\begin{figure*}[!tbh]
\begin{center}
\includegraphics[width=0.5\textwidth]{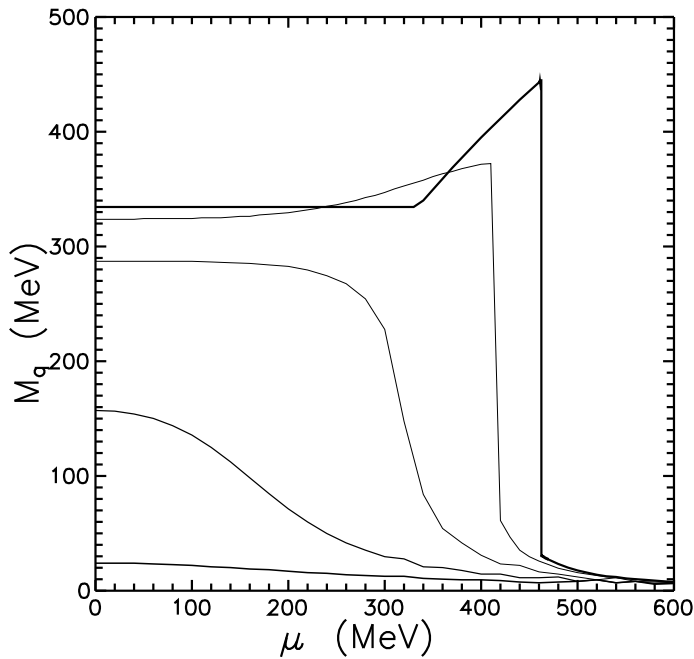}
\end{center}
  \vspace{-7mm}
 \caption{The dynamical quark mass as the function of chemical potential for different temperature
values $T=0$ (upper line), $T=50$, $T=100$, $T=150$, $T=200$ MeV.}
  \label{f5}
\end{figure*}

The Fig. \ref{f1} shows the effective potential $W$ as the quark mass function for the different
values of chemical potential $\mu$. The lower curve corresponds to zero value of chemical potential
and is in full coincidence with the respective curve of the NJL model. With the chemical potential
increasing the process of filling up the Fermi sphere starts and it is easy to see that the effect
of pseudo-particle field screening by the quarks of small masses occurs dominating (see Eq.  (\ref{18a})
and the next one). The screening effect leads to diminishing the absolute value of gluon condensate
and, hence, the effective potential of (anti-)instanton and quark system is increasing. It is distinctly
visible in Fig. \ref{f1} in the region of small quark masses. With the quark mass increasing,
the impact of the filling up process is amplified. Starting on some value of small quark mass
the saturation regime manifest itself and the gluon condensate is suppressed. The plateau
is formed and it is well seen in Fig. \ref{f1}. At the larger values of chemical potential
the chiral symmetry restoration starts and the corresponding curve is shown with the
dashed curve in Fig. \ref{f1}. More detailed analysis allows us to conclude the simplifications
made do not depreciate the qualitative picture of screening.

\begin{figure*}[!tbh]
\begin{center}
\includegraphics[width=0.5\textwidth]{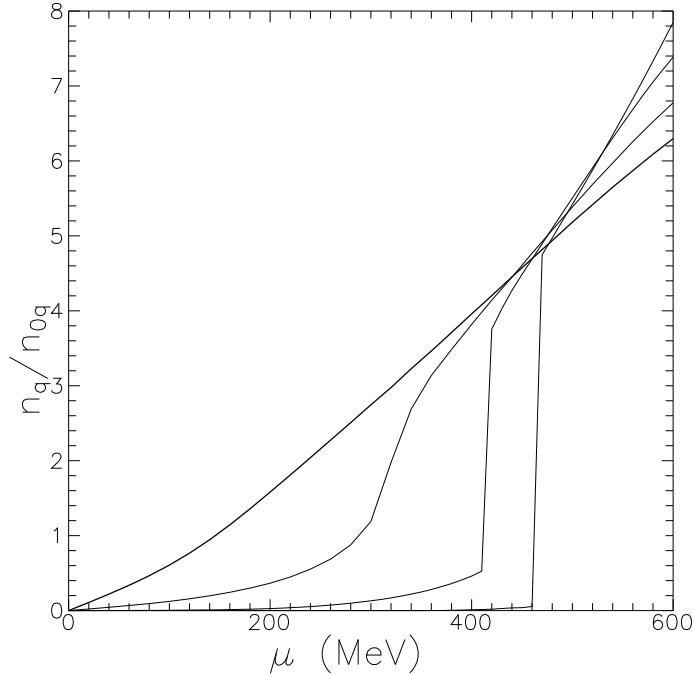}
\end{center}
  \vspace{-7mm}
 \caption{The quark matter density as the function of chemical potential. The last right curve
corresponds to the zero temperature. The next left curve corresponds to the temperature $50$ MeV,
then $100$ MeV, and eventually $T$=$150$ MeV. $n_{0q}=0.062$ fm$^{-3}$ is the normal density of
quark matter calculated from the normal baryon matter density $n_{B}=0.45$ fm$^{-3}$.}
  \label{f6}
\end{figure*}

The IL density as the function of chemical potential at the various values of temperature
(with the temperature increasing the IL density is decreasing) is plotted in the Fig. \ref{f2}.
The dashed curve which was obtained by us in \cite{1}
(the similar mechanism of screening was discussed also in \cite{PY}, \cite{DM88}) corresponds
to the calculation with the massless quarks. The mechanism of forming the observable plateau
is quite understandable. Until the chiral symmetry restoration does not take place the quarks
are ineffecive in the gluon field screening and the gluon condensate practically does not change.

\begin{figure*}[!tbh]
\begin{center}
\includegraphics[width=0.5\textwidth]{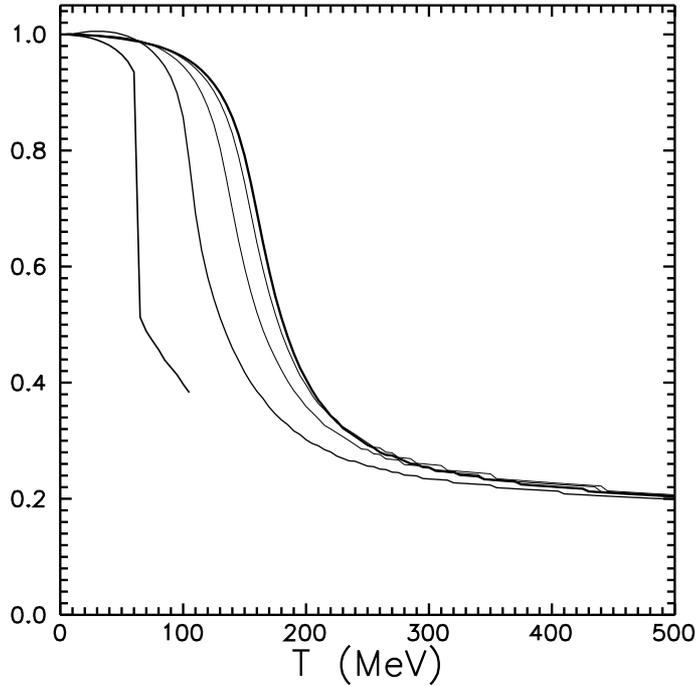}
\end{center}
  \vspace{-7mm}
 \caption{The ratio of quark condensate value to its magnitude at
zero temperature and zero chemical potential as the function of
temperature at the various values of chemical potential
$\mu=0$ (last right curve corresponds to the zero value of chemical
potential), $\mu=110$, $\mu=200$, $\mu=300$, $\mu=400$ MeV (last left curve).}
  \label{f7}
\end{figure*}

Fig. \ref{f3} presents the dependence of IL density on the temperature. As it was expected the
density is larger for the massive quarks than for the massless ones in the region of low
temperature (below $200$ MeV). This result agrees qualitatively with the observation done in \cite{chu}.
Some lattice calculations support this scenario of screening. For example, in \cite{lat} it was
proven that the Debye screening mass behaves as $m_{el}\sim g T$ and depends on the quark flavours
(see Eq. (\ref{8})). Exponential suppression of gluon field with increasing temperature was also found out
in the lattice measurements of correlation functions dealing with the cooled configurations \cite{20}.
In fact, gazing into the detailed analysis of the problem under consideration  we collected a lot of
reasons to have the topological solution with the suppressed chromoelectrical component instead of
the (anti-)coloron one to construct more realistic approach. In particular, it was noticed in \cite{af}
the coloron solution mainly does not fit the lattice data. However, we understand such global pretension
(as a disproving conclusion) is rather naive because our result here shows the coloron solution is quite
practical for estimating the screening effect.

\begin{figure*}[!tbh]
\begin{center}
\includegraphics[width=0.5\textwidth]{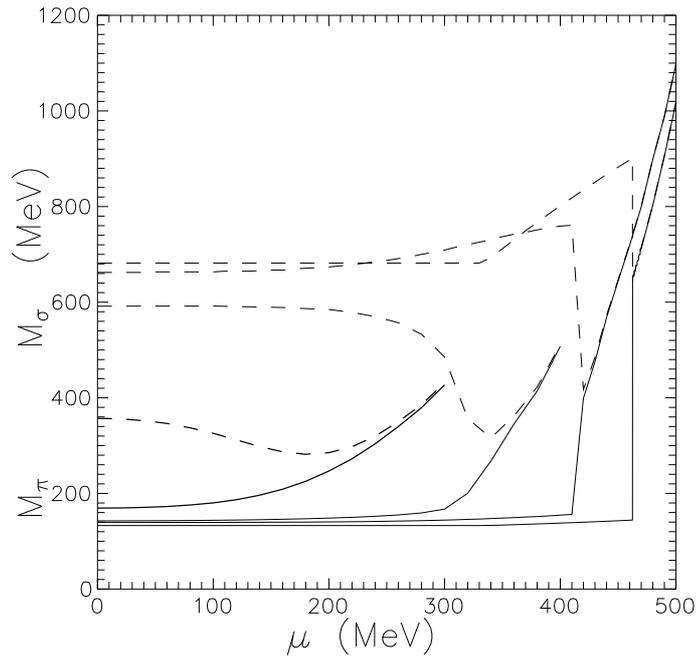}
\end{center}
  \vspace{-7mm}
 \caption{The masses of $\pi$- and $\sigma$-mesons as the functions of chemical potential at different
values of temperature $T=0$, $T=50$, $T=100$, $T=150$ MeV. The dashed lines correspond to the
$\sigma$-meson and the upper dashed line corresponds to the zero temperature. Last right solid line shows
the $\pi$-meson mass behaviour at zero temperature.}
  \label{f8}
\end{figure*}

Analyzing the quark sector we calculated the behaviour of dynamical quark mass as the function
of chemical potential and plotted it in Fig. \ref{f5} for various values of temperature. The upper curve
corresponds to zero temperature and the quark mass behaviour along this curve coincides with the
NJL model up to the chemical potential value of $\mu\sim 300$ MeV. With a further increase of $\mu$
the quark mass increases. It is quite understandable qualitatively if one looks at Fig. \ref{f1}.
At the commencement of chemical potential increase the screening effect does not produce any noticeable
impact on the minimum of effective potential $W$. In spite of effective potential increase in the region
of small quark mass the threshold value of chemical potential should be reached at which the forming
plateau begins to expel the effective potential minimum to the larger mass values. The size of region
in which the quark mass increase takes place is comparable with the quark mass (order of 100 MeV)
and is of interest, in particular, for investigating the equation of the state of strongly interacting matter.
In the version of the NJL model with the parameter choice suggested by T. Hatsuda and T. Kunihiro \cite{NJL}
the chiral symmetry restoration occurs at quite low density (the order of normal nuclear matter density).
In our approach the significant decrease of quark mass is shifted (drags on) to the region of large
chemical potential values approximately 100 MeV larger, which agrees entirely with an intuitive expectation.
Fig. \ref{f6} is devoted to analyzing the quark matter density as the function of chemical potential. It
seems the shift of chiral symmetry restoration phase transition to the region of larger chemical potential
values (order of $400$ MeV), could generate an essential increase of quark matter density. However,
Fig. \ref{f6} demonstrates the change in this interval is inessential (the increase of quark mass
$n_q\sim (\mu^2-M^2)^{3/2}$ provides the compensation), and in actual fact we have to deal with the same
vacuum quarks as at $\mu<300$ MeV.

The quark condensate normalized to its value at zero temperature and zero chemical potential is depicted in
Fig. \ref{f7} as the function of temperature for different values of chemical potential. The behaviours shown
are in full agreement with the predictions of the other models. Finally, two last Figs give more information
on the masses of $\pi$- and $\sigma$-mesons which are also calculated in the NJL model (see, for example,
M.K. Volkov, A.E. Radzhabov \cite{NJL}). The $\pi$-meson mass is given by
\begin{equation}
\label{pi}
M_\pi^{2}=g^2_{\pi qq}~\frat{m}{2G M}~,
\end{equation}
where $g^2_{\pi qq}=\frat{1}{4 I_2}$ is the renormalized constant of meson field interaction including the
following auxiliary integral
$$I_2=\frat{N_c}{8\pi^2}~\int_0^{\widetilde\Lambda} \frat{p^2}{\varepsilon_{p}^3} (1-n_p^{-}-n_p^{+})~dp~.$$
The mass of $\sigma$-meson is defined by the mass of $\pi$-meson and dynamical quark mass as
\begin{equation}
\label{si}
M_\sigma^{2}=M_\pi^{2}+4 M^2~.
\end{equation}
Pion decay constant which is a key element of model tuning is defined as $F_\pi=\frat{M}{g_{\pi qq}}$. Fig. \ref{f8}
presents the masses of $\pi$- and $\sigma$-mesons as the functions of chemical potential for various values of
temperature. The dashed curves correspond to $\sigma$-meson. The upper dashed line shows behaviour at zero
temperature and the solid lower line corresponds to zero temperature behaviour of the $\pi$-meson mass. The interval in
which the $\sigma$- and $\pi$-meson masses become identical defines the parameters (on the $\mu$-$T$ plot)
corresponding to the chiral symmetry restoration. It is clear from Fig. \ref{f8} that such a restoration at zero
temperature occurs around $\mu\simeq 460$ MeV, at $T=50$ MeV around $\mu\simeq 410$ NeV, at $T=100$ MeV around
$\mu\simeq 350$ MeV and at $T=150$ MeV around $\mu\simeq 220$ MeV. Besides, this plot allows us to fix the
line $m_\sigma=2m_\pi$ on which the strong decay channel of $\sigma$-meson is close. At low temperature
($T<100$ MeV) the $\pi$-meson mass undergoes a significant change in the region of chiral symmetry restoration only and
the estimate that the line $m_\sigma=2m_\pi$ approximates chiral symmetry restoration curve ($m_\sigma=m_\pi$)
looks rather practical. At $T=100$ MeV the chemical potential for the line $m_\sigma=2m_\pi$ is about
$\mu\simeq 320$ MeV and at $T=150$ ╠¤┬ we have $\mu\simeq 0$ MeV. The details of this line behaviour could be
quite indicative for searching the mixed phase in relativistic heavy ion collisions \cite{zinsis}.

\section{Conclusion}
In the present paper we investigated the effect of gluon condensate screening with the massive quarks filling up
the Fermi sphere. We developed the approach based on the NJL model highlights which allows us to get
informative qualitative estimates. In particular, we argue that one of the manifestations of filling up the
Fermi sphere could be an increase of the quark mass and, hence, the shift of chiral symmetry restoration
phase transition to the larger values (about $\simeq 100$ MeV) of quark chemical potential. Another instructive
result obtained implies that the gluon condensate does not die out completely in the parameter region characteristic
for this phase transition even at the most advantegeous regime of vacuum gluon field screening. The lattice
measurements of the same quantity confirm such a conclusion allowing us to predict that gluon condensate are
surviving even in the region of parameters essentially beyond the values admissible for our approximation.

The authors are very grateful to I.V. Anikin, M.K. Volkov, S.B. Gerasimov, P. Giubellino, G.V. Efimov,
Yu.B. Ivanov, Yu.L. Kalinovsky, A.E. Kuraev, A.E. Radzhabov, V.V. Skokov, V.D. Toneev,
V.L. Yudichev for numerous fruitful discussions.
Financial support within the Grant INTAS 2004-398 is greatly acknowledged.


\end{document}